\documentclass[aps,twocolumn,showpacs,preprintnumbers,prb,amsmath,amssymb,amsfonts,superscriptaddress,floatfix]{revtex4-1}

\usepackage[latin1]{inputenc}
\usepackage{multirow}
\usepackage{graphics}
\usepackage{color}
\usepackage{amssymb}
\usepackage{amsmath}
\usepackage{overpic}
\usepackage{array}
\usepackage{mciteplus}
\usepackage{bm}
\usepackage{graphicx}
\usepackage{subfigure}

\begin{document}

\title{Competing magnetic orders in quantum critical Sr$_3$Ru$_2$O$_7$}

\author{Aditya Putatunda}
\thanks{These authors contributed equally}
\affiliation{Department of Physics and Astronomy, University of Missouri, Columbia, MO 65211 USA}
\author{Guanhua Qin}
\thanks{These authors contributed equally}
\affiliation{Department of Physics and Astronomy, University of Missouri, Columbia, MO 65211 USA}
\affiliation{Department of Physics, Shanghai University, Shanghai, 200444 China}
\affiliation{Shanghai Key Laboratory of High Temperature Superconductors, MGI and ICQMS, Shanghai University,
Shanghai, 200444 China}
\author{Wei Ren}
\email{renwei@shu.edu.cn}
\affiliation{Department of Physics, Shanghai University, Shanghai, 200444 China}
\affiliation{Shanghai Key Laboratory of High Temperature Superconductors, MGI and ICQMS, Shanghai University,
Shanghai, 200444 China}
\author{David J. Singh}
\email{singhdj@missouri.edu}
\affiliation{Department of Physics and Astronomy, University of Missouri, Columbia, MO 65211 USA}
\affiliation{Department of Chemistry, University of Missouri, Columbia, MO 65211 USA}

\begin{abstract}
We investigated Sr$_3$Ru$_2$O$_7$, a quantum critical metal
that shows a metamagnetic quantum phase transition and electronic nematicity, through density functional calculations.
These predict a ferromagnetic ground state in contrast to the experimentally observed paramagnetism,
raising the question of competing magnetic states and associated fluctuations that may suppress magnetic order.
We did a search to identify such low energy antiferromagnetically ordered metastable states.
We find that the lowest energy antiferromagnetic state has a striped order.
This corresponds to the E-type order that has been shown to be induced by Mn alloying.
We also note significant transport anisotropy in this E-type ordered state.
These results are discussed in relation to experimental observations.
\end{abstract}

\maketitle

\section{Introduction}

Quantum criticality, especially in the context of its material dependent signatures is of significant current interest.
\cite{sachdev_qpt,gegenwart_heavy_fermion_qc}
Here, we investigate the competing orders present in the quantum critical metamagnet Sr$_3$Ru$_2$O$_7$.\cite{grigera329,mm_qc_sr327,iwaya2007local,wu2011quantum}
We find low energy antiferromagnetically ordered states that energetically compete with ferromagnetism.
Interestingly, the lowest energy antiferromagnetic (AFM) states show substantial in-plane transport anisotropy,
which we discuss in relation to nematicity.

Members of the Ruddlesden-Popper (RP) series of strontium ruthenate compounds,
Sr$_{n+1}$Ru$_n$O$_{3n+1}$ have many interesting characteristics.
The $n$=$\infty$ member SrRuO$_3$ is a rare 4\textit{d} itinerant ferromagnet.
\cite{srruo3_fm_singh,srruo3_allen} The $n$=1 member Sr$_2$RuO$_4$, however, is a known unconventional superconductor.
\cite{maeno_94_sc,ishida_98_triplet,2019constraints_nature}
The $n$=2 bilayer compound, Sr$_3$Ru$_2$O$_7$,
the focus of the present work, shows quantum criticality under magnetic field. Its phase diagram shows a
metamagnetic transition with a critical point that
can be tuned to near zero temperature
by applying magnetic field.
\cite{grigera329,lester2015field,grigera2003angular,bruin2013study,puetter2010microscopic}
Borzi and co-workers reported a strong in-plane conductivity anisotropy in this near tetragonal compound
around the critical
point and characterized it as nematic.
\cite{nematicity_review,borzi07}
More broadly, Sr$_3$Ru$_2$O$_7$ presents an interesting case of a nearly ferromagnetic (FM) 4\textit{d}
material with a layered crystal structure and considerable tunability of properties.
\cite{kiyanagi_structure,cava_tetragonal,cao_fm,ikeda_fermi_liquid,ikeda99instability,tipping_fe_doping,afm_order_doping,prl_van_hove_327,problems_dft_fluctuations,chen2016hidden,allan}

In general,
various low-temperature properties of a system situated near a magnetic quantum critical point (QCP),
including transport, are strongly influenced by its associated spin fluctuations,
sometimes up to relatively high temperatures.
\cite{capogna_fluctuations,mqc_o_nmr_327,shaked_distortion,stone2006temperature}
This is the case in Sr$_3$Ru$_2$O$_7$, implying that the spin fluctuations associated with the critical point
are relatively strong in this material.
The underlying quantum fluctuations also lead to a suppression of magnetic order.
\cite{moriya2012spin}
In addition, they also present challenges to the characterization of such systems.
\cite{conundrum}
Commonly employed density functional theory (DFT) approximations,
such as the local density approximation (LDA),
behave like a mean-field theory in this regard and do not capture the effect of such spin fluctuations
that arise near a quantum critical point.
\cite{al_ga_aguayo}
These large fluctuations lead to a systematic overestimation of ground state magnetizations in DFT calculations.
\cite{pd_renormalize,zrzn2_singh}

We note that the overestimation of magnetizations and magnetic moments in standard density functional
calculations for materials is unusual.
In weak and moderately correlated magnetic materials standard DFT yield generally good agreement
with experiment. This includes materials such as the 3d ferromagnets (Fe, Co, Ni and a wide variety of
intermetallics based on them), \cite{gunnarsson,williams,fu}
as well as the ferromagnetic perovskite SrRuO$_3$, \cite{srruo3_fm_singh,mazin-2}
which is chemically and structurally very similar to Sr$_3$Ru$_2$O$_7$.
In strongly correlated systems, such as Mott insulators, the moments are often strongly underestimated
by standard DFT calculations. For example, in the undoped parents of the high temperature cuprate
superconductors, DFT calculations fail to produce the experimentally
observed antiferromagnetic ground states.
\cite{pickett}
In these systems, the Coulomb repulsion, which is needed to localize the electrons, is
inadequately represented in standard DFT calculations. Adding an additional Hubbard $U$
then improves the description, including reproduction of the ground state of undoped cuprates.
\cite{anisimov,dudarev}

While such strongly correlated materials, where standard
DFT calculations underestimate magnetic ordering and do not properly describe the ground state,
are relatively common,
materials where such calculations overestimate the magnetic moments are much less common.
These are cases where spin fluctuations, often associated with nearby quantum critical points,
are strong enough to significantly reduce the bare DFT moments.
This has been discussed in terms of a bare DFT energy surface as a function of magnetization
that is then renormalized by spin fluctuations using a fluctuation amplitude and
a fluctuation renormalized Landau theory analogous to lowest order self-consistent phonon theory.
\cite{shimizu,kaul_renormalize,moriya-book}
Applying this in a quantitative way to predict the renormalized
magnetic properties from first principles is not straightforward due to the difficulty in
determining a cutoff to distinguish spin fluctuations associated with the critical point, not included
in standard DFT, from higher energy spin fluctuations that are included.
\cite{pd_renormalize}
However, by comparing standard DFT calculations with experiment, estimates have been
made of fluctuation amplitudes.
\cite{pd_renormalize,zrzn2_singh}
Not surprisingly, addition of Coulomb correlations by methods such LDA+U degrades agreement
with experiment in these cases since it introduces shifts opposite to those needed.
\cite{johannes-optics}
Furthermore, the magnitude of this type of deviation between DFT and experiment
has been used as a signature to identify materials near magnetic quantum critical points,
\cite{zrzn2_singh,leithe-jasper,krishnamurthy,singh-yfe2si2}
including successful predictions confirmed by
subsequent experiments,
as in the cases of hydrated Na$_x$CoO$_2$ and YFe$_2$Ge$_2$.
\cite{singh-naco2o4,ihara,ihara-2,yfe2ge2-1,yfe2ge2-2,yfe2ge2-3,yfe2ge2-4}

It is also of interest to note the connection of Sr$_3$Ru$_2$O$_7$ and its magnetism
to other members of
the RP series, (Sr,Ca)$_{n+1}$Ru$_n$O$_{3n+1}$.
As mentioned, SrRuO$_3$ is a ferromagnet \cite{kanbayashi} with itinerant character
that is well described by LDA calculations
as far as its magnetism is concerned.
\cite{srruo3_fm_singh,mazin-1,maiti,miao}
Furthermore, details of its electronic structure, including for example, LDA based predictions
of a negative spin polarization have been confirmed in detail by experiments.
\cite{nadgorny,raychaudhuri,worledge}

Theoretical work indicates significant sensitivity of the magnetism to structure in this
compound. \cite{mazin-1,rondinelli,zayak}
Experimentally, alloying with Ca leads to increased distortion of the ideal perovskite structure
through octahedral tilts.
This is accompanied by a decrease in the magnetic ordering temperature until a critical point
is reached at $\sim$70\% Ca, beyond which a highly renormalized near ferromagnetic metal
is found.
\cite{cao-ca,kikugawa}

The importance of octahedral tilts and rotation in relation to magnetism is also found
in single layer (Sr,Ca)$_2$RuO$_4$.
Sr$_2$RuO$_4$ is a paramagnetic Fermi liquid, that exhibits unconventional superconductivity
at low temperature. \cite{maeno_94_sc,ishida_98_triplet,2019constraints_nature}
There has been debate about the extent and nature of correlations in this material.
\cite{pchelkina,singh-comment,ingle,damascelli}
However, it is generally agreed that the Fermi surface agrees
with that predicted by LDA calculations,
\cite{oguchi,singh-95}
although with mass renormalization,
\cite{katsufuji,bergemann}
that spin-fluctuations likely play an important role in the superconductivity
\cite{sigrist,mazin-2,mazin1999competitions}
and that these spin fluctuations have a substantial itinerant origin.
This itinerant behavior includes the observation of incommensurate spin fluctuations
predicted on the basis of Fermi surface nesting.
\cite{braden}
Alloying with Ca in (Sr,Ca)$_2$RuO$_4$ again demonstrates sensitivity to structure.
Initially there is an increasing ferromagnetic susceptibility as the octahedra rotate, 
followed by a crossover, and eventually near pure Ca$_2$RuO$_4$ the development of an antiferromagnetic
insulating phase with a strong change in the Ru-O bond lengths
reflecting distortion of the octahedra. \cite{nakatsuji-1,carlo}

In any case, the fluctuation-dissipation theorem, which relates the
amplitude of the fluctuations to the dissipation, given by an integral
involving the imaginary part of the susceptibility, implies an enhanced imaginary component of
the magnetic susceptibility associated with the sizable fluctuations in materials near
magnetic quantum critical points.
This in turn points towards the presence of strongly competing orders in materials
that show strong spin fluctuations but no order, as discussed previously.
\cite{kaul_renormalize}
Besides an overly strong tendency towards ferromagnetism,
both FM and AFM fluctuations
\cite{capogna_fluctuations,perry2000hall}
may coexist in this ruthenate system
\cite{mqc_o_nmr_327,mazin1999competitions}
Here we report a search for such competing orders including commonly
discussed AFM states
as well as the so called E-type order that occurs with heavy Mn doping.
\cite{mesa_e_type}

\section{Computational Methods}

We searched initially for various possible magnetic orders using projector augmented wave
(PAW) pseudopotentials as implemented in Vienna Ab-initio Simulation Package (VASP)
\cite{vasp1,vasp2}.
An energy cutoff of 500 eV was used.
Energy and force convergence criteria were chosen as 10$^{-7}$ eV and 0.01 eV\AA$^{-1}$, respectively.
The Brillouin zone (BZ), in this case, was sampled on a 5$\times$5$\times$5 \textbf{k}-mesh.
We used both LDA and the Perdew-Burke-Ernzerhof (PBE) generalized gradient approximation (GGA).
\cite{gga_made_simple}
We also checked the structural predictions of these two functionals.
We find that the PBE functional leads to a unit cell volume 1.6\%
larger than experiment (average lattice parameter error of +0.5\%),
while the LDA leads to an underestimate of the unit cell volume by 6.6\%
(average lattice parameter error of -2.2\%).
These are within the range of typical errors for these functionals and the somewhat smaller
lattice parameters predicted by LDA relative to PBE is also as usual.

Following this survey,
we then investigated the low lying states in detail using the
general potential linearized augmented plane wave (LAPW) \cite{singh2006planewaves} method
as implemented in the WIEN2k code.
\cite{blaha2001wien2k}
The LAPW sphere radii for Ru and Sr atoms were both chosen as R$_{MT}$=2.1 Bohr,
while for O atoms R$_{MT}$=1.55 Bohr was used.
The basis size was set by plane-wave cutoff K$_{max}$ with R$_{min}$K$_{max}$=7.0.
This leads to an effective RK$_{max}$=9.5 for the metal atoms.
The self-consistent calculations were performed using a BZ sampling of
at least 1000 \textbf{k}-points in the respective BZs.
Transport integrals were done using the BoltzTraP code.
\cite{boltztrap}
Dense Brillouin zone sampling with \textbf{k}-meshes of dimensions 30$\times$16$\times$16 or higher was
used for these calculations.

\begin{figure}[tbp]
\centering
\includegraphics[width=\columnwidth]{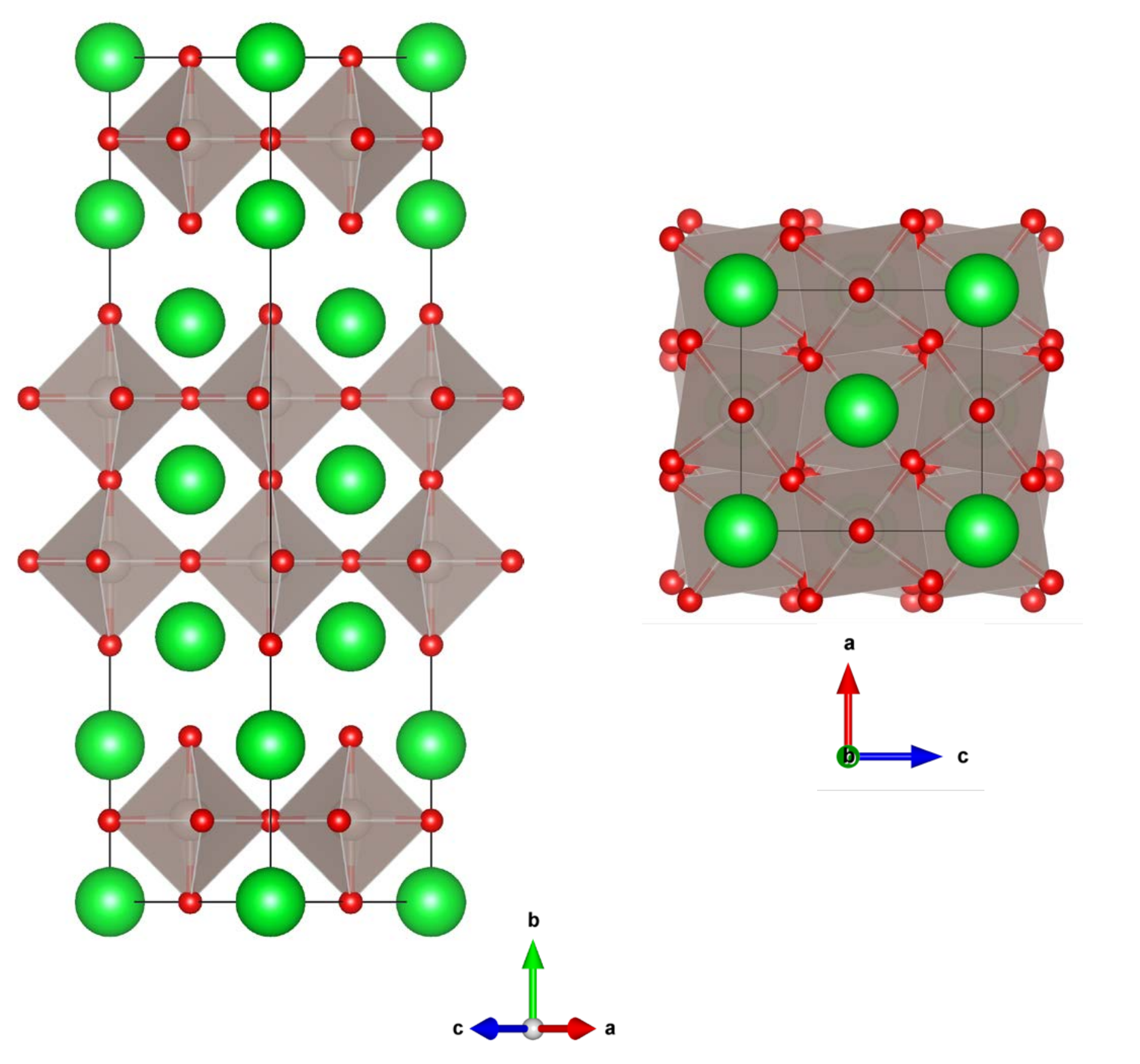}
	\caption{Crystal structure of orthorhombic Sr$_3$Ru$_2$O$_7$, showing layering along $c$ (left) and view
	along the $c$-axis, illustrating octahedral rotations (right).}
\label{structure}
\end{figure}

Sr$_3$Ru$_2$O$_7$ has a layered perovskite structure
that is formed by two sheets of corner sharing RuO$_6$ octahedra connected via a shared apical oxygen
(Figure \ref{structure}).
Interestingly, the metal ions still occupy the ideal tetragonal symmetry
sites similar to the n=1 compound Sr$_2$RuO$_4$, although the Ru-O-Ru bonds are bent due to the counter-rotation
of the octahedra about the $c$-axis.
These rotations amount to approximately 7$^{\circ}$ and are opposite for the two sheets making up a bilayer.
This in combination with the stacking of the bi-layers
reduces the overall symmetry so that finally the compound has an orthorhombic crystal structure,
space group Ccca (\#68).
\cite{shaked_distortion,uniaxial_press,cava_mag_order,hu2010surface}

The lattice parameters for our calculations were taken from experiment,
specifically the measurements performed on single crystals,
as reported by Kiyanagi and co-workers
\cite{kiyanagi_structure}
These are \textit{a}=5.4979 \AA, \textit{b}=5.5008 \AA, and \textit{c}=26.7327 \AA.
The internal positions of the atoms were relaxed. Details of the structure are in the supplemental material.
\cite{supplemental}

\section{Results and discussion}

\subsection{Magnetic Order}

Sr$_3$Ru$_2$O$_7$
is a known metal and despite having a susceptibility peak at around $\sim$18 K,
\cite{cava_mag_order,ikeda99instability}
it displays no long range magnetic order.
\cite{ikeda_fermi_liquid}
Multiple reports however, suggest temperature-dependent competing FM and AFM spin fluctuations,
although the nature of the AFM fluctuations has not been established.
\cite{perry_prl_2001,capogna_fluctuations,ti_doped_supress,problems_dft_fluctuations,lda_slave_boson,multiorbital_phys,ruthenates_hetero_struc}

Both experimental and theoretical investigations show that the material lies close to a magnetic instability.
\cite{ikeda_fermi_liquid,tight_binding_instability,tipping_fe_doping}
Apart from applying external magnetic field along various directions,
investigations by perturbing the system via uniaxial pressure,
\cite{uniaxial_press,hidden_phases_stress,rivero17uniaxial_tilt}
doping by both magnetic impurities\cite{afm_order_doping,tipping_fe_doping}
and non-magnetic impurities\cite{ti_doped_supress,ti_doped_sdw}
find various magnetic behaviors.
DFT investigations find a ferromagnetic instability in contrast to the experimentally observed enhanced
paramagnetic state.
\cite{ikeda_fermi_liquid,electronic_struc_singh,jpsj_327_theory}
As mentioned this type of error is a characteristic of a material near a quantum critical point.
In the case of Sr$_3$Ru$_2$O$_7$, the predicted ferromagnetism has an itinerant origin, coming from
a high density of states associated with the structure of the $t_{2g}$ bands, specifically van Hove
singularities in the $d_{xy}$ band.
This Stoner mechanism is similar to SrRuO$_3$ and Sr$_2$RuO$_4$.
\cite{prl_van_hove_327,electronic_struc_singh,mazin-1,mazin-2,lee}
The finding of an incorrect ferromagnetic state is not affected by spin orbit coupling.
In our calculations, which were done in a scalar relativistic approximation,
we found a spin magnetization of 4.7 $\mu_B$ per unit cell (4 Ru atoms, including the interstitial
and O components) in the LDA, which is
reduced by less than 10\% to 4.4 $\mu_B$ per cell when spin orbit is included.
Adding Coulomb correlations using the LDA+U method, \cite{dudarev}
with a moderate value $u$=4 eV and the standard fully localized limit double counting
strongly increases the magnetization to 7.9 $\mu_B$ per cell, opposite to what is needed to
produce agreement with experiment.
This is not unexpected, since
degradation of standard DFT results with the addition of $U$ has been noted in other
itinerant magnetic systems previously.
\cite{cococcioni,fu}

As mentioned, Sr$_3$Ru$_2$O$_7$ displays a
metamagnetic transition at a field strength of approximately 7 to 8 T.\cite{mm_qc_sr327}
However, doping using magnetic impurities
\cite{hossain2013electronic} such as Fe\cite{tipping_fe_doping} and Mn
\cite{mesa_e_type,afm_order_doping,chen2016hidden,hossain2008crystal,li2013atomic}
as well as non-magnetic Ti
\cite{ti_doped_sdw} 
has been found to yield different AFM orders.
\cite{hossain2013electronic}
In general, the relationship between these and properties of the the undoped compound
is unclear, since these dopants produce strong perturbations of the system.
Nonetheless, one notable order is the double stripe E-type order that is observed with heavy Mn doping,
\cite{mesa_e_type} although it is reported to have a short correlation length.
\cite{mesa_e_type,tipping_fe_doping}
It is to be noted that in particular it is quite
unclear that the the E-type order found in Mn doped samples is
reflective of the properties of undoped Sr$_3$Ru$_2$O$_7$.
This is because the Mn doping is also
accompanied by sizable
distortions in the crystal structure along with
band width changes that may stabilize antiferromagnetic structures.
\cite{sr327_phase_dia,autieri}
Furthermore, doping in Sr$_3$Ru$_2$O$_7$
is often accompanied with transition to a state of more insulating transport,
\cite{mesa_e_type,afm_order_doping,tipping_fe_doping,ti_doped_sdw,another_e_type_exp,sr327_phase_dia}
while pristine Sr$_3$Ru$_2$O$_7$ is clearly metallic and itinerant.
This has led to a focus on other orders as possible competing orders to ferromagnetism
in pristine Sr$_3$Ru$_2$O$_7$.
For example, based on their hybrid functional calculations,
Rivero and coworkers reported other AFM orders both of metallic and insulating nature
that may be obtained via strain, particularly a layered A-type AFM.
\cite{half_metallic_rivero,hidden_phases_stress,brodsky2017strain,marshall2018electron}

\begin{figure}[tbp]
\centering
\includegraphics[width=\columnwidth]{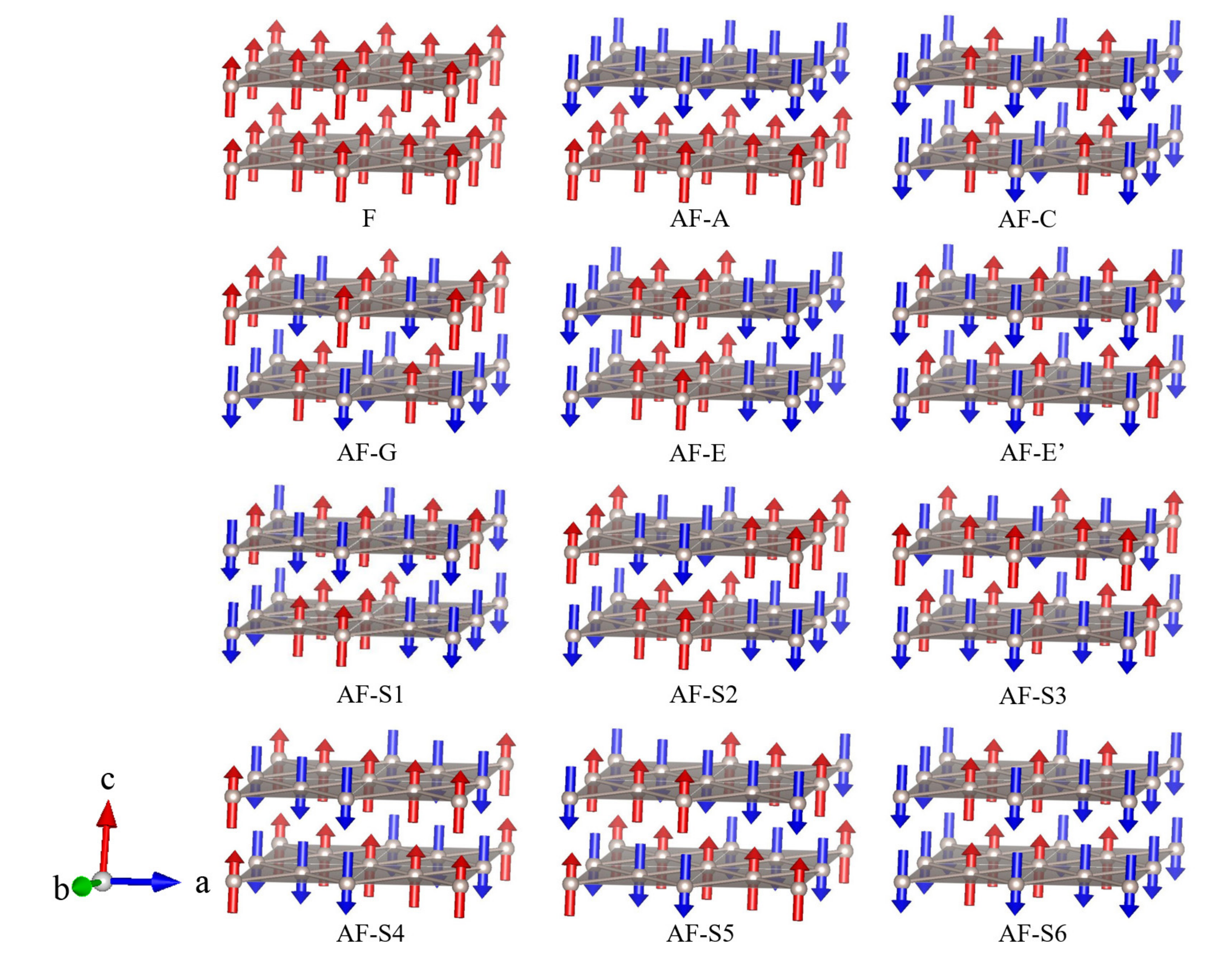}
\caption{Various magnetic configurations investigated. Only Ru atoms are shown. These include ferromagnetism,
	A-type, G-type, and C-type antiferromagnetism, which are common orders for perovskites,
	two E-type orders, which are slightly non-degenerate due to the orthorhombic crystal symmetry,
	and more complex orders with larger unit cells.}
\label{convention}
\end{figure}

Here we performed a search for possible long range AFM orders
(within collinear magnetism) in relation to both FM and non-magnetic orders
initially through PAW calculations (Figure \ref{convention}).
We find the FM state as the ground state for both LDA and PBE functionals.
Besides the FM order, the lowest energies are for the E-type order.
There are two such states that are slightly non-degenerate due to orthorhombicity.
PBE shows stronger magnetism over LDA including larger magnetic energies and higher moments.
The three other commonly discussed AFM states lie much higher in energy in the order,
A$<$C$<$G (Figure \ref{convention}). No self-consistent solution was found for the G-type order within the LDA.
The bottom six S-labeled AFM orders lie somewhere in the middle of the whole range.
Details of the magnetic energies as obtained from these initial PAW calculations are in
the Supplemental Material (SI).
\cite{supplemental}

The lack of solution for G-type and the variability of the moments between the different states
is a characteristic of itinerant magnetism,
as is the fact that the energy differences between different orders
are of similar order to the energy difference between the non-spin polarized and the lowest energy FM state.
Thus local moment pictures, such as short range Heisenberg models, are not well suited to this material.
Furthermore, the A-type order, which consists of oppositely aligned
FM layers stacked across the bilayer, lies much higher in energy than both of the E-types and the FM order.
This means that interactions between the layers within a bilayer are strong.

We now turn to the detailed results obtained with the LAPW method.
\cite{problems_dft_fluctuations,singh2006planewaves} 
The energetics and magnetic moments are in Table \ref{wien_energetics}.
Most importantly, we find that the two metastable E-type AFM states carry large magnetic moments
($\sim$1.08 $\mu_B$ within PBE and $\sim$0.85 $\mu_B$ within LDA).
As expected, these are the orders which consistently lie closest to the FM ground state.

\begin{table}[h]
\caption{Energy ordering of various magnetic orders (per formula unit) found by LAPW calculations and
their (absolute) averaged  magnetic moments for both PBE and LDA functionals.
Refer to Figure \ref{convention} for naming. FM and NM respectively stand for ferromagnetic and nonmagnetic state (zero energy level), while all the other orders are antiferromagnetic in nature. Note that the moments reported here are those lying within the LAPW sphere radii.}
\vspace{0.01\textheight}
\begin{tabular}{c cc cc}
\hline
\hline
\multirow{2}{*}{} & \multicolumn{2}{c}{PBE}                                                                                                                                            & \multicolumn{2}{c}{LDA}                                                                                                                                            \\ \cline{1-5} 
          Order        & \multicolumn{1}{c}{\begin{tabular}[c]{@{}c@{}}$\Delta$E\\ meV/f.u.\end{tabular}} & \begin{tabular}[c]{@{}c@{}}Mag mom\\ Ru ($\mu_B$)\end{tabular} & \multicolumn{1}{c}{\begin{tabular}[c]{@{}c@{}}$\Delta$E\\ meV/f.u.\end{tabular}} & \begin{tabular}[c]{@{}c@{}}Mag mom\\ Ru ($\mu_B$)\end{tabular} \\ \hline
FM                     & -147.8 & 1.28                                                     & -29.9  & 0.73                                                     \\
E                  & -131.5  & 1.08                                                     & -28.6  & 0.85 \\
E$'$                  & -130.3  & 1.08                                                     & -28.0  & 0.85                                                     \\
A                  & -95.6  & 1.06                                                     & -21.2  & 0.60                                                     \\
C                  & -43.6 & 0.72                                                     & -13.6  & 0.38                                                     \\
G                  & -3.2 &  0.45                                                     & --     & --                                                      \\
NM                     & 0       & 0.000                                                         & 0       & 0.000                                                     \\
\hline
\hline
\end{tabular}
\label{wien_energetics}
\end{table}

The sizable moments obtained and the FM ground state contradict the fact that
Sr$_3$Ru$_2$O$_7$ is an experimentally determined paramagnet.
However, this is almost certainly due to the fact that the system lies close to
a magnetic QCP where strong spin fluctuations suppress any long-range magnetic order in the system.
Standard DFT calculations fail to describe this type of fluctuations,
as has been noted for other such materials near a magnetic QCPs.
\cite{problems_dft_fluctuations,zrzn2_singh,subedi_nbfe2,qcp_srrho3_singh,ni_al_ga_prl_qcp}
As mentioned, this overestimation of magnetism has been used as a signature of materials that are in
the vicinity of a QCP.
\cite{pd_renormalize,zrzn2_singh,kaul_renormalize,ni_al_ga_prl_qcp}

In addition, one may note that the magnetic moments predicted for the E-type orders
are indeed the largest among the AFM states.
All the other investigated orders lie higher in energy and have lower magnetic moments.
The energy difference within LDA between the FM and E orders is only 
1.6 meV per formula unit on average (for the E and E').
Thus we find that the E-type order is very likely the order that competes with ferromagnetism in this material.
It is interesting to note that the E-type order
is also the order among the ones considered that breaks the tetragonal symmetry within the RuO$_2$ plane.

\subsection{Electronic structure and transport}

\begin{figure}[tbp]
\centering
\includegraphics[width=0.9\columnwidth]{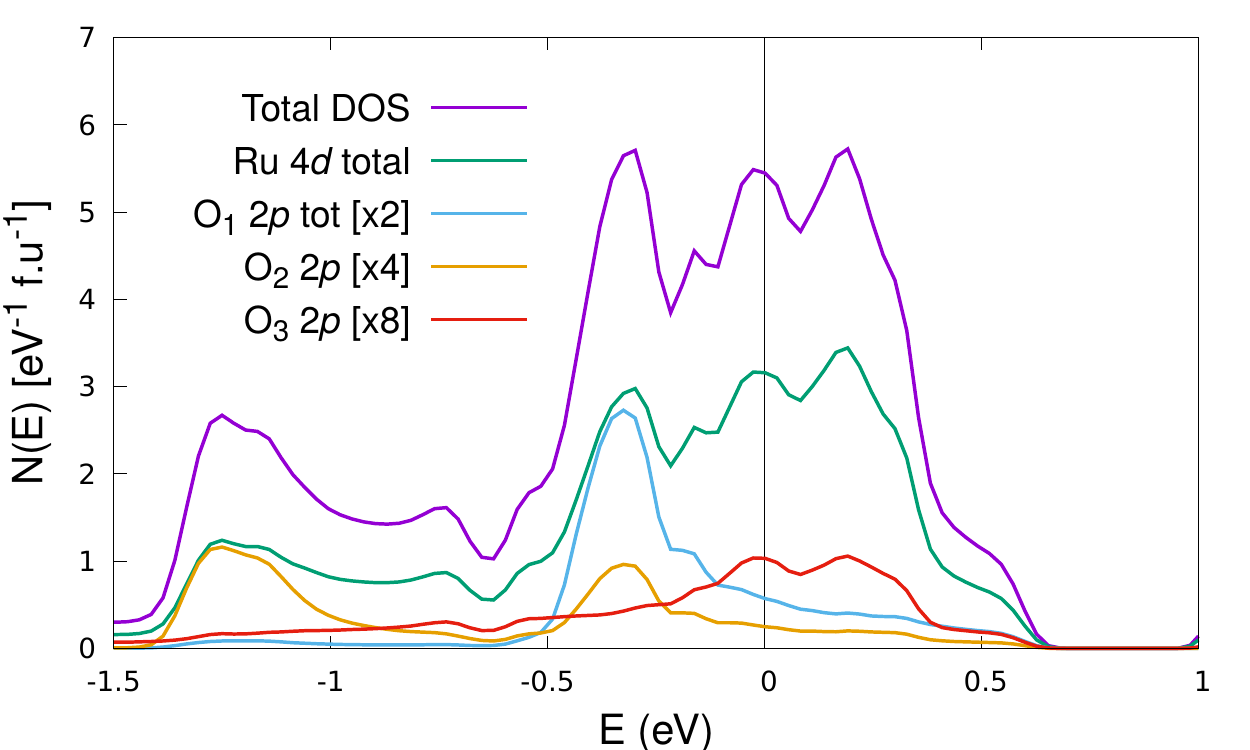}
\caption{Density of states per formula unit of Sr$_3$Ru$_2$O$_7$ for its nonmagnetic
state showing the respective LAPW sphere projected contributions
from Ru-4\textit{d} and O-2\textit{p} orbitals obtained within LDA.
Three kinds of O contributions (scaled 4x for visibility) are observed (see text)
with their respective multiplicities, shown in parentheses. The black vertical line at E=0 shows the Fermi level.}
\label{dos}
\end{figure}

Our calculated electronic density of states (DOS) of non-magnetic Sr$_3$Ru$_2$O$_7$,
as shown in Figure \ref{dos} shows a peak around the Fermi level. This
favors magnetism through the Stoner mechanism as discussed previously.
\cite{mazin-1}
For the considered E-type orders,
the corresponding DOS (shown in SI)\cite{supplemental}
is distorted but still high near the Fermi level.
Table \ref{dos_ef} summarizes the DOS value observed at the Fermi level, N(E$_F$)
for each investigated magnetic order.
Table \ref{dos_ef} shows that the electronic structure remains metallic
for all the spin-orderings considered. As noted previously,
there is strong Ru 4\textit{d}-O 2\textit{p} hybridization evident.

\begin{table}[h]
\caption{Density of states (per formula unit) at the Fermi level, N(E$_F$) of Sr$_3$Ru$_2$O$_7$ for various magnetic orders with both LDA and PBE functionals. Note that for the ferromagnetic order (FM), and for other magnetic orders, N(E$_F$) for each single spin channel ($\uparrow$,$\downarrow$) is shown. A Gaussian broadening of 4 meV was used.}
\vspace{0.01\textheight}
\begin{tabular}{c cccc} 
\hline
\hline
\multirow{2}{*}{Order} & \multicolumn{4}{c}{LAPW}                   \\
                       & \multicolumn{2}{c}{LDA} & \multicolumn{2}{c}{PBE}  \\ 
\hline
FM ($\uparrow$,$\downarrow$)                     & 3.9 & 3.4               & 0.9 & 4.8                 \\
E                      & \multicolumn{2}{c}{5.0} & \multicolumn{2}{c}{4.5}  \\
E'                     & \multicolumn{2}{c}{5.4} & \multicolumn{2}{c}{4.7}  \\
A                      & \multicolumn{2}{c}{4.0} & \multicolumn{2}{c}{3.0}  \\
C                      & \multicolumn{2}{c}{7.8} & \multicolumn{2}{c}{7.1}  \\
NM                     & \multicolumn{2}{c}{5.4} & \multicolumn{2}{c}{6.4}  \\
\hline
\hline
\end{tabular}
\label{dos_ef}
\end{table}

\begin{figure*}[tb]
\includegraphics[width=\textwidth]{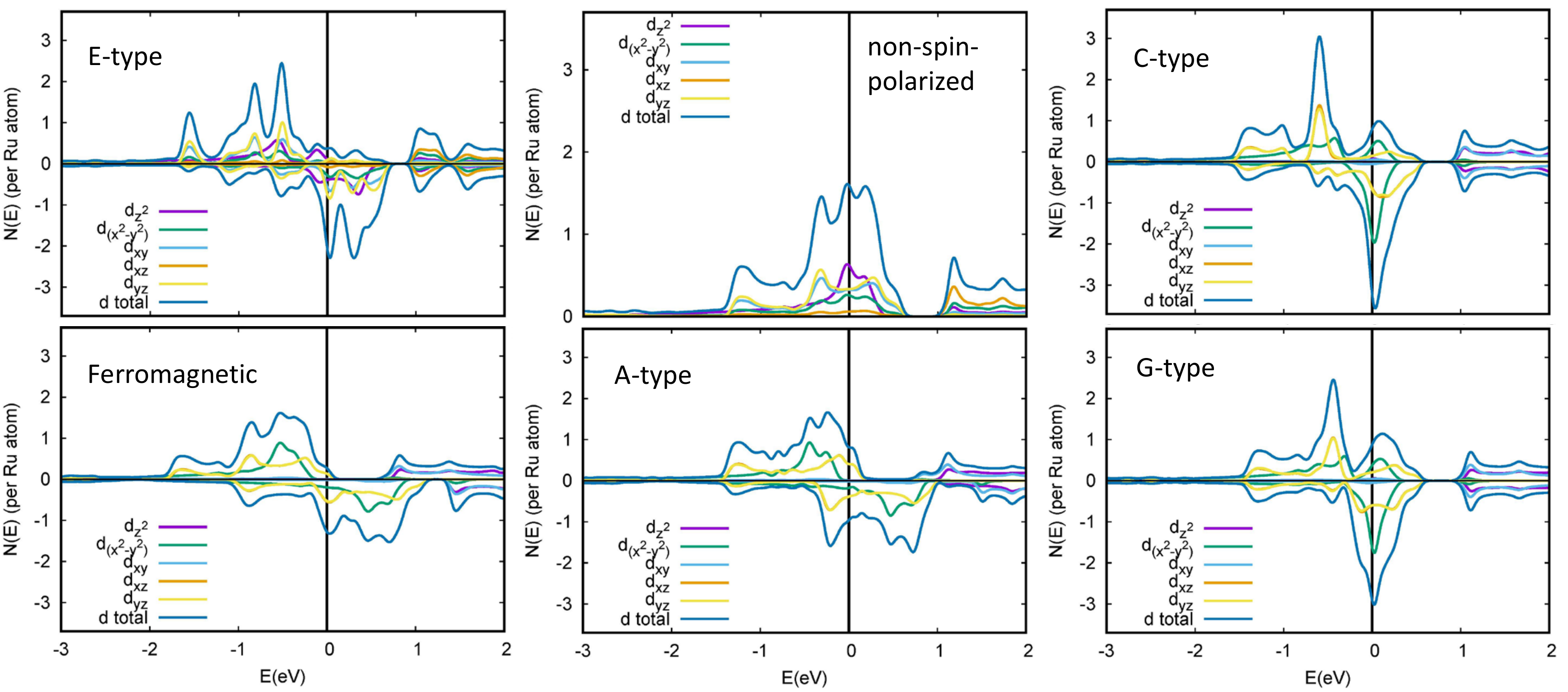}
\caption{Projected DOS of $d$ character on a Ru atom in the energy
range of the $t_{2g}$ bands for the different magnetic orderings as obtained
with the PBE functional. Note that the individual $d$ orbitals are mixed because of the low symmetry
induced by the octahedral rotations and magnetic order. The same symmetry was used for the E-type and
non-spin-polarized. The ferromagnetic, A-type, C-type and G-type were done in a smaller higher symmetry
cell which leads to a different coordinate system for the $d$ orbitals.}
\label{newdos}
\end{figure*}

The individual contributions from each of the three different types of O atoms are labeled.
It can be noted that the 2\textit{p} contribution from the O3 atoms, which are the in-plane O,
is the largest in the region closest around the Fermi level.
It reaffirms the fact that the material is highly two dimensional
and most of the electronic transport occurs primarily in-plane.
O1 and O2 are respectively the shared and SrO layer apical oxygen atoms and contribute less near the Fermi level.

In an octahedral crystal field, the $d$ orbitals split into a lower lying $t_{2g}$ manifold, with
three bands (and six electrons) and a higher lying $e_g$ manifold that can accommodate four electrons.
The $e_g$ manifold is derived from $\sigma$ antibonding combinations of O $p$ and Ru $d$ orbitals, while
the normally more narrow $t_{2g}$ manifold consists of more weakly antibonding $\pi$ combinations.
Ru$^{4+}$, as in Sr$_3$Ru$_2$O$_7$, has four $d$ electrons, which leads to a partially filled $t_{2g}$
manifold that is responsible for the magnetism and transport.
The electronic DOS in the region near the Fermi level is derived from hybridized Ru $t_{2g}$ and O $p$ states.

The orbital character is often important in understanding magnetic ordering, especially
in systems where transition metal - O hybridization is important, for example double exchange systems.
\cite{brzezicki}
Fig. \ref{newdos} shows the projections of Ru $d$ onto a site with the different magnetic orders as
obtained with the PBE functional.
As noted previously, non-spin-polarized Sr$_3$Ru$_2$O$_7$ has a relatively narrow set of nominally
$t_{2g}$ orbitals. \cite{electronic_struc_singh}
It should be noted, however, that this crystal field notation is not strictly correct
since the octahedral rotation mixes the $e_g$ and $t_{2g}$ manifolds, and the layered structure splits the
$t_{2g}$ orbitals. There is also mixing due to symmetry lowering associated with magnetic order
as well as splitting due to interactions between the two layers forming a bilayer.

However, we find that the general shape of the DOS in the energy range of the $t_{2g}$ orbitals does not depend
strongly on magnetic order, showing a higher peak around the Fermi level against a broader peak at $\sim$ -1 eV.
The main effect of magnetism is to exchange split this peak into a lower lying majority and higher lying
minority components, with the largest exchange splitting for the orders where the moment is highest.
The second aspect to note is that the E-type order gives a strong narrowing of the individual DOS peaks
in the $t_{2g}$ manifold. This leads to a greater differentiation of the orbitals. This is also the case for the
C-type and G-type orders, which have nearest neighbor antiferromagnetism in a single plane. Meanwhile
the lowest energy ferromagnetic and the A-type order have generally broader individual peaks.

In Tables \ref{lda-aniso} and \ref{pbe-aniso},
we show the reduced electrical conductivity ($\sigma/\tau$) values for different orders
obtained using both LDA and PBE functionals.
The transport integrals were done for a temperature of 100 K in the Fermi function for
computational convenience.
These were calculated using the BoltzTraP code.\cite{boltztrap}
The BoltzTraP code constructs a smooth interpolation of the energy bands that passes through
all the first principles points.
In our calculations we used dense first principles meshes consisting of
30$\times$16$\times$16 grids or better so that the interpolated bands are accurate.
The BoltzTraP code then does transport
integrals using this interpolation to construct the gradients that comprise the band velocities.
To ensure consistency across various magnetic orders,
the conductivity tensors have been appropriately diagonalized and only the in-plane directions are given.
These are the two largest eigenvalues of the conductivity tensor.
In Tables \ref{lda-aniso} and \ref{pbe-aniso},
the out-of-plane reduced conductivity components,
being about 2 orders smaller then the in-plane components have been omitted.

\begin{table}[tbp]
\centering
\caption{In-plane components of the diagonalized reduced electrical conductivity tensor
	and the corresponding anisotropies for various magnetic orders with the LDA functional.}
\begin{tabular}[c]{c >{\centering}m{6em}>{\centering}m{6em} c} 
\hline
\hline
	Order     & \multicolumn{2}{c}{\begin{tabular}[c]{@{}c@{}}$\sigma/\tau$ (10$^{18}$ $(\Omega ms)^{-1}$) \end{tabular}} & \begin{tabular}[c]{@{}c@{}}In-plane \\ anisotropy \end{tabular}  \\ 
\hline
E     &  65  & 77                                                                                                                                                 & 1.17                                                               \\
E'     &  67  & 75                                                                                                                                                 & 1.13                                                               \\
A     &  249 & 250                                                                                                                                                & 1.00                                                               \\
\multirow{2}{*}{FM} ($\uparrow$) &  159 & 161                                                                                                                                                & 1.01                                                               \\
\hspace{18pt}($\downarrow$)      &  191 & 191                                                                                                                                                & 1.00                                                               \\
C     &  229 & 233                                                                                                                                                & 1.02                                                               \\
NM    &  266 & 274                                                                                                                                                & 1.03                                                               \\
\hline
\hline
\end{tabular}
\label{lda-aniso}
\end{table}

\begin{table}[tbp]
\caption{In plane components of the diagonalized reduced electrical conductivity tensor and their corresponding anisotropies for various magnetic orders with the PBE functional.}
\begin{tabular}[c]{c >{\centering}m{6em}>{\centering}m{6em} c}
\hline
\hline
	Order     & \multicolumn{2}{c}{\begin{tabular}[c]{@{}c@{}}$\sigma/\tau$\ (10$^{18}$ $(\Omega ms)^{-1}$) \end{tabular}} & \begin{tabular}[c]{@{}c@{}}In plane \\ anisotropies \end{tabular}  \\
\hline
     E     &  27                  &          49              &  1.77 \\
     E'    &  24                  &          47              &  1.95 \\
     A     &  174                 &          177             &  1.01 \\
\multirow{2}{*}{FM} ($\uparrow$)  &  11.4 &  11.5  &  1.01                                                     \\
\hspace{18pt}($\downarrow$)       &  218 & 222 &  1.02                                                     \\
    C     &  187                  &          189             &  1.01  \\
    G     &  248                  &          256             &   1.03  \\
    NM  &   256  &          264            &  1.03  \\
\hline
\hline
\end{tabular}
\label{pbe-aniso}
\end{table}

As seen in Tables \ref{lda-aniso} and \ref{pbe-aniso},
noticeable anisotropy occurs among the in-plane conductivity components only in case of the E-type magnetic order.
These anisotropy values are noticeably larger than the ones obtained for any other orders.
One may note that within LDA, the in-plane (reduced) electrical conductivity values differ
by about $\sim$15\%.
While it is perhaps not surprising that the E-type order gives more anisotropy considering that
the pattern of magnetic moments in the RuO$_2$ planes is anisotropic with this order, unlike other
simple orders, it is important that this anisotropy in the magnetic pattern is indeed well reflected
in the electronic structure at the Fermi level that controls transport.
The higher-symmetry (and lower energy)
E-type order is slightly more anisotropic than the E' order.
In PBE, however, the anisotropies are larger.
This reflects its tendency towards larger moments.
In this case, when contrasted to LDA, the ordering is reversed and E' order has higher anisotropy.

It is interesting to note that in their investigation Borzi and co-workers\cite{borzi07}
found an in-plane resistivity anisotropy value of $\sim$20\%
near their lowest reported experimental temperature.
This is within the range of the conductivity anisotropies found for the E-type orders, for example, $\sim$15\%
on average for the LDA and the larger values for the PBE functional, which has larger moments.

\section{Summary and Conclusions}

We investigated Sr$_3$Ru$_2$O$_7$,
a quantum critical material to identify its low-lying antiferromagnetic metastable states
that might contribute to the strong spin fluctuations that are thought to strongly affect its properties.
We find that the energetically lowest metastable states carry a striped E-type AFM ordering.
The corresponding transport properties show sizable in-plane conductivity anisotropy
in contrast to other possible AFM orders. Experimental investigation
using inelastic neutron scattering in search
of spin-fluctuations arising from this E-type order will be of interest.

\acknowledgments

Work at the University of Missouri is supported by the Department of Energy,
Basic Energy Sciences award DE-SC0019114.
GQ is grateful for support from the China Scholarship Council (CSC).
Support for work at Shanghai University
is provided by the National Natural Science Foundation of China
(Grants 51672171, 51861145315 and 51911530124),
Independent Research Project of State Key Laboratory of Advanced Special Steel and Shanghai Key
Laboratory of Advanced Ferrometallurgy at Shanghai University, and the fund of the State Key Laboratory
of Solidification Processing in NWPU (Grant SKLSP201703).

\bibliography{refs}
\end{document}